\documentclass[twocolumn,english,prl,superscriptaddress]{revtex4}
\usepackage[T1]{fontenc}
\usepackage[latin9]{inputenc}
\usepackage{amsmath}
\usepackage{amssymb}
\usepackage{graphicx}
\usepackage{esint}

\makeatletter
\@ifundefined{textcolor}{}
{%
 \definecolor{BLACK}{gray}{0}
 \definecolor{WHITE}{gray}{1}
 \definecolor{RED}{rgb}{1,0,0}
 \definecolor{GREEN}{rgb}{0,1,0}
 \definecolor{BLUE}{rgb}{0,0,1}
 \definecolor{CYAN}{cmyk}{1,0,0,0}
 \definecolor{MAGENTA}{cmyk}{0,1,0,0}
 \definecolor{YELLOW}{cmyk}{0,0,1,0}
}

\newcommand{\ex}{{\mathbf e}_x}                            % e_x
\newcommand{\ey}{{\mathbf e}_y}                            % e_y
\newcommand{\ez}{{\mathbf e}_z}                            % e_z

\usepackage{amsfonts}%\usepackage{newlfont}
\usepackage{subfigure}\usepackage{lscape}

\usepackage{babel}

\makeatother

\usepackage{babel}
\begin{document}

\title{Magnetically generated spin-orbit coupling for ultracold atoms}

\author{Brandon M. Anderson}

\email{brandona@umd.edu}

\affiliation{Joint Quantum Institute, University of Maryland, College Park, Maryland
20742-4111, USA 20742, USA}

\affiliation{National Institute of Standards and Technology, Gaithersburg, Maryland
20899, USA }

\author{I. B. Spielman}

\affiliation{Joint Quantum Institute, University of Maryland, College Park, Maryland
20742-4111, USA 20742, USA}

\affiliation{National Institute of Standards and Technology, Gaithersburg, Maryland
20899, USA }

\author{Gediminas Juzeli\={u}nas}

\affiliation{Institute of Theoretical Physics and Astronomy, Vilnius University,
A. Go\v{s}tauto 12, Vilnius 01108, Lithuania}
\begin{abstract}
We present a new technique for producing two and three dimensional
Rashba-type spin-orbit coupling for ultra cold atoms without involving
light. The method relies on a sequence of pulsed inhomogeneous magnetic
fields imprinting suitable phase gradients on the atoms. For sufficiently
short pulse durations, the time-averaged Hamiltonian well approximates
the Rashba Hamiltonian. Higher order corrections to the energy spectrum
are calculated exactly for spin-$1/2$ and perturbatively for higher
spins. The pulse sequence does not modify the form of rotationally
symmetric atom-atom interactions. Finally, we present a straightforward
implementation of this pulse sequence on an atom-chip.
\end{abstract}
\maketitle
Proposals for creating Rashba type spin-orbit coupling (SOC) in cold
atomic gases abound \cite{Dudarev2004,Ruseckas2005,Stanescu2007a,Jacob2007,Juzeliunas2008PRA,Stanescu2008,Vaishnav08PRL,Juzeliunas2010,Campbell2011,Dalibard2011,Zhai2012JMPB,Xu2012,Anderson2012PRL,Galitski2013}.
All of these schemes rely (at least partially \cite{Xu2012}) on the
coupling of atoms to laser beams. Unfortunately, the atom-light interaction
is associated with spontaneous emission, leading to heating or loss.
To date, only the experimentally most simple case -- an equal mixture
of Rashba and Dresselhaus SOC -- has been realized in the lab \cite{Lin2011,Chen2012,Wang2012,Cheuk2012}.
Implementation of Rashba SOC would allow for the study of rich ground
state physics proposed in systems of many-body fermions \cite{Liu2009,Vyasanakere:2011,Vyasanakere:2011bis,Jiang2011,Hu:2011,Liu:2012,Chen2012PRA,He2012a,Cui2013,Zheng2013,Mundo2013}
and bosons \cite{Stanescu2008,Sinha2011,Radic2011PRA,Wu11CPL,Hui2012,Gou2012,Kawakami2012,Ruokokosk2012,Xu12PRA,Sedrakyan2012PRA,Song2012arXiv},
of which many properties have no condensed matter analogue.

Rashba SOC can be intuitively understood as a momentum-dependent magnetic
field that is symmetric under simultaneous spin and momentum rotations.
Generically, realizing such behavior requires terms in the atomic
Hamiltonian which link spin to momentum. Laser beams using two-photon
Raman transitions are an obvious choice for implementing such coupling,
as they induce transitions between two internal states while simultaneously
imparting momentum.

Here we demonstrate that Rashba or Dresselhaus SOC can be created
in cold atoms without any optical fields by imprinting phase gradients
in different directions using a properly chosen pulse sequence of
inhomogeneous magnetic fields. This linearly varying magnetic field
provides a uniform spin-dependent force, imparting a desired momentum.
When the direction of the magnetic field and the gradient of its magnitude
are perpendicular to each other, the form of the momentum boost represents
a position dependent rotation of the atomic spin. This suggests pulsed
magnetic field gradients have the necessary features to produce Rashba
or Dresselhaus SOCs. 

The current scheme can be realized in a straightforward manner on
state-of-the-art atom chips \cite{Trinker2008,Goldman:2010}with SOC
strengths comparable with those in optical implementations. In contrast,
the optical schemes rely on using many laser beams that couple internal
atomic states in a complex way. The proposal is applicable to any
atomic species containing an arbitrary non-zero spin and does not
alter the form of $SU(2)$ invariant atom-atom interactions \cite{Ohmi1998,Ho1998,Kawaguchi2012}.
Our proposal allows for study of spin-1 and spin-2 SOC bosons, where
the symmetry of the atom-atom interactions strongly affects the symmetry
of the many-body ground state \cite{Gou2012,Ruokokosk2012,Xu12PRA,Song2012arXiv}.
In optical setups, \cite{Dalibard2011} the adiabatic elimination
of a number of atomic states makes the atom-atom interaction position-dependent
and not $SU(2)$ invariant.

Time averaged descriptions of periodically driven systems can often
acquire gauge fields, the most simple example of which is the transformation
into a rotating frame \cite{Franzosi2004,Fetter2009,Cooper2008}.
One can also generate artificial magnetic fields by combining lattice
and time-dependent quadrupolar potentials \cite{Sorensen2005}, or
by shaking \cite{Graham1992,Madison1998PRL,Eckardt:2005,Eckardt2010,Hemmerich2010,Kolovsky2011,Struck:2012,Arimondo2012}
or stirring \cite{Kitagawa2010} optical lattices. Here we focus on
a different scenario where a time dependent magnetic field yields
SOC (rather than an Abelian magnetic flux) for atoms in the continuum
(rather than on a lattice). Unlike the case for conventional magnetic
trapping where the atomic spin adiabatically follows the local magnetic
field \cite{Migdall1985}, here the field pulses time average to zero,
provide no trapping potential, and lead to dynamic spin evolution. 

\begin{figure}[!t]
\includegraphics[width=1\columnwidth]{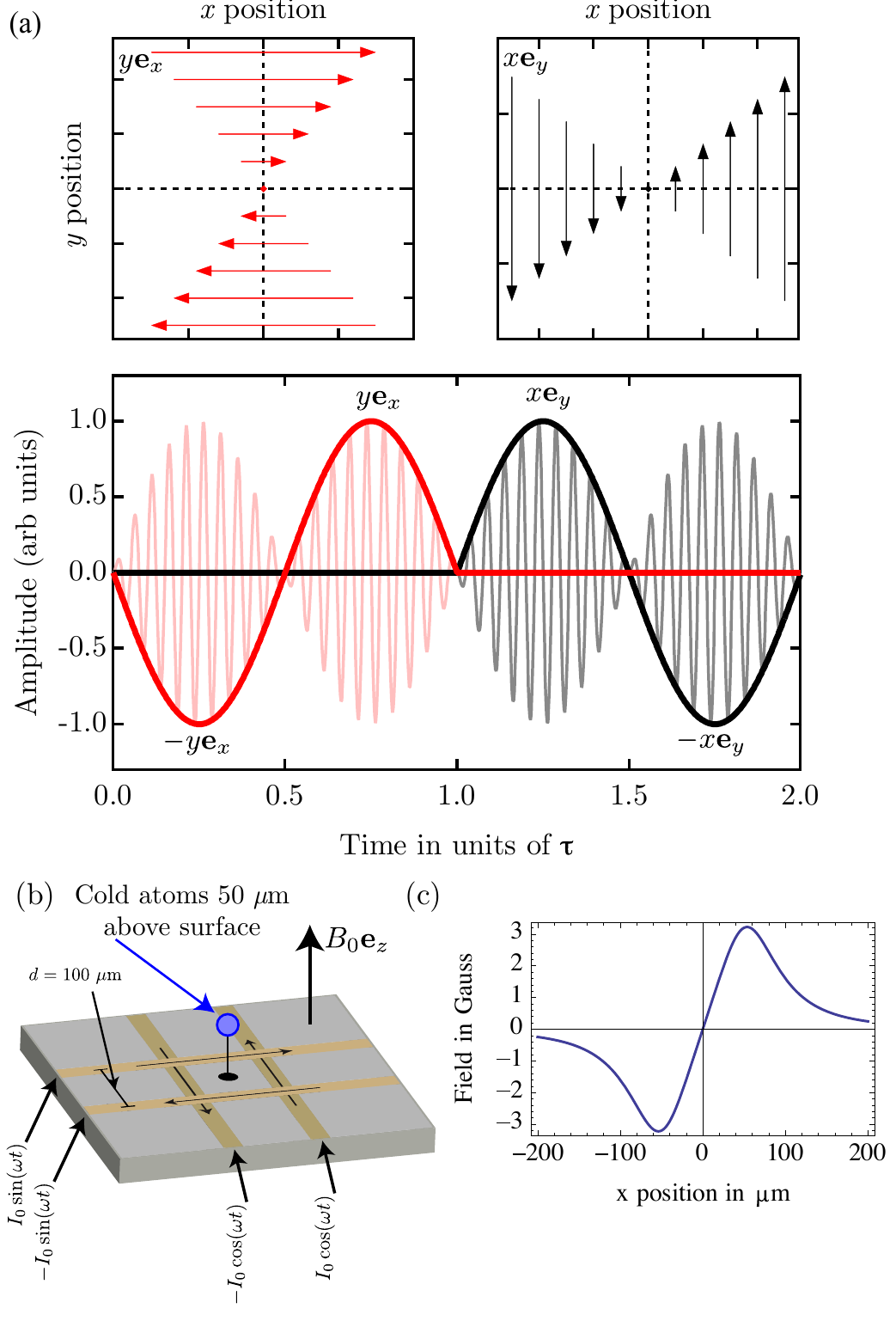} \caption{\label{fig:Experimental-scheme}Proposed atom chip implementation
of 2D Rashba SOC using pulsed magnetic fields. (a) One full pulse
of period $2\tau$. For $0\leq t<\tau$ an effective coupling vector
$\boldsymbol{\Omega}_{x}=-\beta\left(t\right)k_{{\rm eff}}y\ex$ (red)
writes a spin-dependent phase gradient along $y$ in the quantization
basis of $F_{x}$. In the second half of the pulse, $\tau\leq t<2\tau$,
the coupling vector $\boldsymbol{\Omega}_{y}=\beta\left(t-\tau\right)k_{{\rm eff}}x\ey$
(black) writes a phase gradient along $x$ in the quantization basis
of $F_{y}$. (b) The cloud of atoms sits $50\mu{\rm m}$ above the
surface of an atom chip. A strong bias field $B_{0}\ez$ sets a quantization
axis. Two sets of microwires wires parallel to $\ex$ and $\ey$ are
spaced $50\mu{\rm m}$ from the center of the cloud in the $\ex-\ey$
plane carry an amplitude modulated rf current. (c) The current configuration
produces a magnetic field gradient that is linear near the center
of the atom-chip. }
\end{figure}

\paragraph*{General Formulation:}

We focus on the atoms in a spin-$f$ hyperfine ground state manifold
characterized by the spin vector $\mathbf{F}$ with components obeying
the commutation relations $[F_{i},F_{j}]=i\hbar\epsilon_{ijk}F_{k}$.
The interaction of the atom and the magnetic field $\mathbf{B}\equiv\mathbf{B}(\mathbf{r},t)$
is given by the Hamiltonian 
\begin{equation}
H_{\mathbf{B}}(\mathbf{r},t)=\boldsymbol{\Omega}(\mathbf{r},t)\cdot\mathbf{F}\,,\label{eq:H_B}
\end{equation}
with $\boldsymbol{\Omega}\left(\mathbf{r},t\right)=g_{F}\mu_{B}\mathbf{B}$,
where $\mu_{B}$ is the Bohr magneton and $g_{F}$ is the the Land\'{e}
g-factor. The Schr\"{o}dinger equation describing the combined internal
and center of mass evolution of the atom is 
\begin{equation}
i\hbar\partial_{t}\left|\psi\right\rangle =\left[H_{0}+V+H_{\mathbf{B}}(\mathbf{r},t)\right]\left|\psi\right\rangle ,\label{eq:Shroed-Eq}
\end{equation}
with $H_{0}=\mathbf{p}^{2}/2m$, where $\mathbf{r}$ and $\mathbf{p}$
are, respectively, the atomic center of mass coordinate and momentum
operators obeying $\left[r_{i},p_{j}\right]=i\hbar\delta_{ij}$. In
what follows, we will neglect the state-independent trapping potential
$V$.

\paragraph*{1D SOC:}

We first show a properly chosen pulsed magnetic field gradient can
give rise to a 1D SOC. In the first stage, $0\le t<\tau$, an effective
coupling vector $\boldsymbol{\Omega}_{x}=-\beta\left(t\right)k_{{\rm eff}}y\ex$
writes a spin-dependent phase gradient along $\ey$ (where $\mathbf{e}_{xyz}$
denote the Cartesian unit vectors) in the quantization basis of $F_{x}$,
where the wavevector $k_{{\rm eff}}$ characterizes the strength of
the magnetic field gradient, and $\beta\left(t\right)$ defines its
temporal shape. While a magnetic field $\mathbf{B}\sim y\ex$, cannot
exist in region of zero electric currents, in the experimental section
we will show how to produce a coupling Hamiltonian Eq. (\ref{eq:H_B})
corresponding to $\boldsymbol{\Omega}_{x}$ using a strong bias magnetic
field along $\mathbf{e}_{z}$, and a fast oscillating magnetic field
in the $\ex-\ez$ plane, as shown in Fig. \ref{fig:Experimental-scheme}.

To elucidate the main idea, suppose that at times $t=0$ and $t=\tau$,\textbf{
$\mathbf{B}$} is pulsed for a short enough duration that the atoms
hardly move, i.e., $\beta\left(t\right)=\delta\left(t-\epsilon\right)-\delta\left(t-\tau+\epsilon\right)$
with $\epsilon\rightarrow0$. The pulse at $t=0$ rotates the spin
about $\ex$ according to the operator $R_{x}=\exp\left[ik_{{\rm eff}}yF_{x}/\hbar\right]$.
The particle then evolves freely for a time $\tau$ before a second
pulse ``undoes'' the rotation, described by $R_{x}^{\dagger}$.
The total evolution of the particle after both pulses is described
by
\begin{equation}
U_{x}\left(\tau\right)=R_{x}e^{-\frac{iH_{0}\tau}{\hbar}}R_{x}^{\dagger}=\exp\left[-i\frac{\left(\mathbf{p}-k_{{\rm eff}}F_{x}\ey\right)^{2}}{2m\hbar}\tau\right],\,\label{eq:U_1}
\end{equation}
 representing the evolution for a particle with SOC along $\ey$.

The analysis leading Eq. (\ref{eq:U_1}) can be readily extended to
any pulse of finite width and zero average $\int_{0}^{\tau}\beta\left(t\right)dt=0$.
This coupling can be eliminated from Eq. (\ref{eq:Shroed-Eq}) by
the unitary transformation $R_{x}(t)=\exp\left[iF_{x}k_{{\rm eff}}y\int_{0}^{t}\beta\left(t^{\prime}\right)\, dt^{\prime}/\hbar\right]$
which also changes the momentum $\mathbf{p}$ to $\mathbf{p}^{\prime}=\mathbf{p}-k_{{\rm eff}}F_{x}\ey\int_{0}^{t}\beta\left(t^{\prime}\right)\, dt^{\prime}$
in the transformed Hamiltonian $\tilde{H}_{0}\left(t\right)=R_{x}\left(t\right)H_{0}R_{x}^{\dagger}\left(t\right)=\mathbf{p}^{\prime2}/2m$.
The latter $\tilde{H}_{0}\left(t\right)$ commutes with itself at
different times. Thus, using $R_{1}\left(\tau\right)=1\!\!1$, where
$1\!\!1$ is the identity operator in spin space, one can exactly
calculate the time evolution operator $U_{x}\left(\tau\right)=\exp\left[-i\int_{0}^{\tau}\tilde{H}_{0}\left(t\right)\, dt/\hbar\right]$
after one pulse, giving: 
\begin{eqnarray}
U_{x}\left(\tau\right) & = & \exp\left[-\frac{i\tau}{\hbar}\left(\frac{\left(\mathbf{p}-c_{1}k_{{\rm eff}}F_{x}\ey\right)^{2}}{2m}+s\frac{k_{{\rm eff}}^{2}}{2m}F_{x}^{2}\right)\right]\,\label{eq:U_1-exact}
\end{eqnarray}
where $s=c_{2}-c_{1}^{2},$ with $c_{n}=\tau^{-1}\int_{0}^{\tau}\mathrm{d}t^{\prime}\left[\int_{0}^{t^{\prime}}\beta\left(t^{\prime\prime}\right)\mathrm{d}t^{\prime\prime}\right]^{n}$.
For two delta pulses, $\beta\left(t\right)=\delta\left(t-\epsilon\right)-\delta\left(t-\tau+\epsilon\right)$,
we have $c_{1}=1$ and $s=0$, so one arrives at Eq. (\ref{eq:U_1}).
For a smoothly alternating coupling, $\beta\left(t\right)=\frac{2\pi}{\tau}\sin\left(2\pi\frac{t}{\tau}+\varphi\right)$,
where $\varphi$ sets the origin of time, one has that $c_{1}=\cos\varphi$
and $s=1/2$. If the pulse is repeated many times, the choice of $\varphi$
cannot matter. Indeed, the vector potential $c_{1}k_{{\rm eff}}F_{x}\ey$
in Eq. (\ref{eq:U_1-exact}) can be eliminated by a gauge transformation.
In the next section, a second pulse at times $\tau\leq t<2\tau$ breaks
time translation symmetry. The vector potential then cannot be gauged
away, and only signals $\beta_{a}\left(t\right)=\left(\beta\left(t\right)-\beta\left(\tau-t\right)\right)/2$,
antisymmetric over one period, will contribute to $c_{1}$, with a
maximum of $c_{1}=1$ for $\varphi=0$. Since the magnetic pulse strength
depends only on the product $\beta\left(t\right)k_{{\rm eff}}$, we
will henceforth set $c_{1}=1$, which can always be done through a
re-definition of $k_{{\rm eff}}$.

\paragraph*{2D SOC:}

To add SOC in another direction, at $\tau\leq t<2\tau$, we introduce
a second stage during which $\boldsymbol{\Omega}\equiv\boldsymbol{\Omega}_{y}=\beta\left(t-\tau\right)k_{{\rm eff}}x\ey$
with the same temporal behavior as the first stage. The unitary operator
$U_{y}(\tau)$, evolving the system from $t=\tau$ to $t=2\tau$,
has the form of Eq. (\ref{eq:U_1-exact}) with $\ey\rightarrow-\ex$
and $F_{x}\rightarrow F_{y}$. 

These two stages together, with $U_{xy}\equiv U_{y}U_{x}$ describe
a single full cycle of our repeating pulse sequence. Thus the Hamiltonian
is time-periodic with period $2\tau$. For a sufficiently short period,
the combined evolution operator $U_{xy}$ can be well approximated
using the Baker-Campbell-Hausdorff formula, up to first order in $\tau$,
giving $U_{xy}\approx\exp\left[-iH_{2D}2\tau/\hbar\right]$, where
\begin{equation}
H_{2D}=\frac{p^{2}}{2m}-k_{{\rm eff}}\frac{p_{y}F_{x}-p_{x}F_{y}}{2m}+c_{2}k_{{\rm eff}}^{2}\frac{F_{x}^{2}+F_{y}^{2}}{4m}\label{eq:H_F-2}
\end{equation}
 is the effective Hamiltonian describing the evolution of the system
under our repeated pulse sequence, observed at integer multiples of
$2\tau$. 

For the spin-$1/2$ case with $\mathbf{F}=\hbar\boldsymbol{\sigma}/2$,
Eq. (\ref{eq:H_F-2}) reduces to the Rashba Hamiltonian, 
\begin{equation}
H_{2D}=H_{R}=\frac{\mathbf{p}^{2}}{2m}-v\boldsymbol{\sigma}\cdot\left(\ez\times\mathbf{p}\right)\,,\quad\textrm{with}\quad v=\frac{\hbar k_{{\rm eff}}}{4m}\,,\label{eq:Rashba-2D}
\end{equation}
where an overall energy offset has been omitted. For higher spins
($f>1/2$) the last term in Eq. (\ref{eq:H_F-2}) is proportional
to $\hbar^{2}f\left(f+1\right)1\!\!1-F_{z}^{2}$, introducing an effective
quadratic Zeeman (QZ) shift. Since $c_{2}>0$, using the two-dimensional
setup it is thus impossible to completely eliminate the QZ term in
Eq. (\ref{eq:H_F-2}) for $f\neq1/2$, and produce the Rashba Hamiltonian
in Eq. (\ref{eq:Rashba-2D}) with $\boldsymbol{\sigma}$ replaced
by $2\mathbf{F}/\hbar$. The QZ term preserves the conserved quantum
number $J_{z}=L_{z}+F_{z}$, where $L_{z}=xp_{y}-yp_{x}$, so it is
unlikely to significantly affect the ground state phases explored
in systems with higher spin Rashba SOC \cite{Gou2012,Xu12PRA,Ruokokosk2012}. 

We validated this approach by numerically simulating the trapped,
weakly interacting Gross-Pitaevskii equation and non-interacting Schr\"{o}dinger
equation. We applied the periodic pulse sequence described above to
the ground state without SOC, and slowly ramped on $k_{{\rm eff}}$.
When measured at full periods, the system was similar to the ground
states found in Ref. \cite{Sinha2011,Hui2012}. We also used imaginary
time propagation to find the true Rashba ground state, followed by
the pulsed Rashba SOC described above, where $k_{{\rm eff}}$ was
matched to the minimum of the Rashba ring. If viewed at complete periods,
the system did not significantly deviate from the many-body ground
state. These numerical tests suggest that the proposed SOC system
well approximates Rashba SOC.

\begin{figure}[!t]
\includegraphics[width=1\columnwidth]{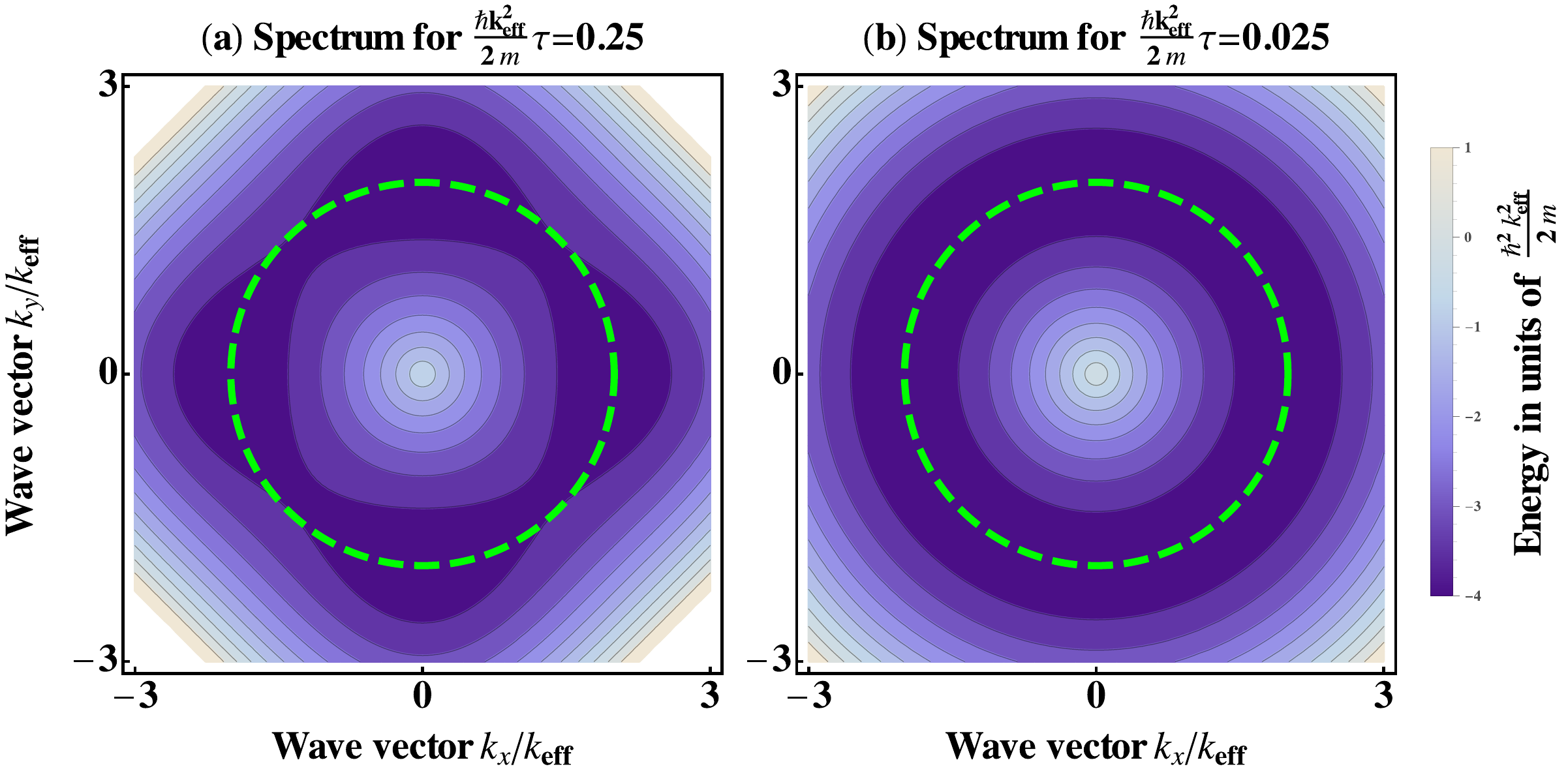} 

\caption{\label{fig:E-exact} The lower band of the the exact time-averaged
Hamiltonian of a spin-$1/2$ particle under the influence of the pulsed
magnetic field scheme described in the main text. (a) For long evolution
times, $\hbar k_{{\rm eff}}^{2}\tau/2m=0.25$, the spectrum has a
Bloch structure of periodicity $4\pi m/k_{{\rm eff}}\tau$ superposed
on the $\mathbf{p}^{2}/2m$ dispersion of a free particle. (b) Short
evolution times, $\hbar k_{{\rm eff}}^{2}\tau/2m=.025$ well approximate
the Rashba ring. The green dashed line represents the minimum energy
``Rashba ring'' of $H_{2D}$ in Eq. (\ref{eq:Rashba-2D}).}
\end{figure}

\paragraph*{3D Spin-orbit coupling}

Pulsed magnetic fields can provide not only the conventional two-dimensional
Rashba or Dresselhaus coupling, but also three-dimensional (3D) SOC
which is not encountered for electrons in condensed matter structures.
By adding an additional pulse oriented along the $\ez$ direction,
all three spin matrices $\left\{ F_{x},F_{y},F_{z}\right\} $ can
be coupled to momentum. For instance, using three magnetic coupling
stages $\boldsymbol{\Omega}\equiv\boldsymbol{\Omega}_{x}=-\beta\left(t\right)k_{{\rm eff}}x\ey$
for $0\le t<\tau$, $\boldsymbol{\Omega}\equiv\boldsymbol{\Omega}_{y}=-\beta\left(t-\tau\right)k_{{\rm eff}}y\ez$
for $\tau\le t<2\tau$ and $\boldsymbol{\Omega}\equiv\boldsymbol{\Omega}_{z}=-\beta\left(t-2\tau\right)k_{{\rm eff}}z\ex$
for $2\tau\le t<3\tau$ and using a procedure analogous to the one
presented above, one can simulate a spin-orbit coupling 
\begin{equation}
H_{3D}=\frac{\mathbf{p}^{2}}{2m}-v\frac{2}{\hbar}\left(p_{x}F_{y}+p_{y}F_{z}+p_{z}F_{x}\right)\,,\label{eq:Rashba-3D}
\end{equation}
with $v=\hbar k_{{\rm eff}}/6m$, describing the 3D SOC of the Rashba
type. This scheme has an additional advantage over 2D Rashba spin-orbit
coupling in that the quadratic term featured in Eq. (\ref{eq:H_F-2})
is proportional to $\mathbf{F}^{2}=\hbar^{2}f\left(f+1\right)1\!\!1$,
which is a constant and thus has been omitted in Eq. (\ref{eq:Rashba-3D}).
Therefore, the 3D setup simulates a pure spin-orbit coupling without
a quadratic Zeeman term for arbitrary spin systems. The present proposal
allows for creating 3D SOC in a more simple manner without any use
of optical fields. Previous proposals to produce a 3D SOC involve
complex optical transitions between four or more internal atomic states
\cite{Anderson2012PRL,Li2012}.

\paragraph*{Corrections to the 2D Single particle spectrum:}

In the derivation of Eq. (\ref{eq:H_F-2}), the product $U_{y}U_{x}$
was expanded to lowest order in the short time $\tau$. In practice,
a finite pulse time $\tau$ will result in deviations from the ideal
SOC form. For a spin-$1/2$ system, the effective Hamiltonian for
the 2D system can be calculated exactly using $U_{y}U_{x}=\exp\left(-i\mathbf{p}^{2}\tau/m\hbar\right)S_{\left(yx\right)}^{\dagger}S_{\left(xy\right)}$,
where $S_{\left(uv\right)}=\exp\left(i\gamma_{u}\sigma_{v}\right)=\cos\gamma_{u}+i\sigma_{v}\sin\gamma_{u}$
is a rotation matrix and $\gamma_{u}=k_{{\rm eff}}p_{u}\tau/2m$ is
the corresponding momentum-dependent angle. The product of two rotations
is itself a rotation $S_{\left(yx\right)}^{\dagger}S_{\left(xy\right)}=\exp\left(i\gamma\hat{\mathbf{n}}\cdot\boldsymbol{\sigma}\right)$
around an axis $\hat{\mathbf{n}}$ by an angle $\gamma\equiv\gamma\left(\mathbf{p}\right)$,
implicitly defined by $\cos\gamma=\cos\gamma_{x}\cos\gamma_{y}$ and
$\hat{\mathbf{n}}\sin\gamma=\ex\cos\gamma_{x}\sin\gamma_{y}-\ey\sin\gamma_{x}\cos\gamma_{y}-\ez\sin\gamma_{x}\sin\gamma_{y}$.
The exact time-averaged Hamiltonian is then given by $H_{2D}^{\left(\mathrm{exact}\right)}=\frac{\mathbf{p}^{2}}{2m}-\frac{\hbar}{2\tau}\gamma\hat{\mathbf{n}}\cdot\boldsymbol{\sigma}+{\rm const}$. 

This allows for a straightforward calculation of the deviations of
the time-averaged eigenstates from the ideal Rashba form. The lower
band has energy given by $E\left(\mathbf{p}\right)=\frac{\mathbf{p}^{2}}{2m}-\frac{\hbar}{2\tau}\gamma\left(\mathbf{p}\right)$.
We plot this spectrum as a function of momentum in Fig. (\ref{fig:E-exact}).
For long pulses, $\tau\gg2m/\hbar k_{{\rm eff}}^{2}$, the structure
resembles a periodic band structure with an overall $\mathbf{p}^{2}/2m$
envelope. The periodicity of $\gamma\left(\mathbf{p}\right)$ in momentum
space is given by $\hbar k_{p}=4\pi m/k_{{\rm eff}}\tau$. In the
opposite limit where $\tau\ll2m/\hbar k_{{\rm eff}}^{2}$, the periodicity
of $\gamma\left(\mathbf{p}\right)$ becomes much longer than the characteristic
momentum $\hbar k_{{\rm eff}}$ which sets the momentum scale of the
Rashba spin-orbit term. Thus, for sufficiently short pulses and sufficiently
small momentum, the spectrum well approximates Rashba spin-orbit coupling. 

For spin-$f>1/2$ particles, the QZ terms $F_{x}^{2}$ and $F_{y}^{2}$
featured in $U_{x}$ and $U_{y}$ do not allow for an exact solution.
For short pulses, $\hbar k_{{\rm eff}}^{2}\tau/2m\ll1$, higher order
corrections to the average Hamiltonian can be found perturbatively.
The first order correction $\delta\bar{H}^{\left(1\right)}$ is 
\begin{eqnarray}
\delta\bar{H}^{\left(1\right)} & = & \frac{\tau}{4}\left(\frac{k_{{\rm eff}}^{2}}{2m}\right)^{2}\left(-4\frac{p_{x}p_{y}}{k_{{\rm eff}}^{2}}F_{z}-2c_{2}\frac{p_{y}}{k_{{\rm eff}}}\left\{ F_{y},F_{z}\right\} \right.\nonumber \\
 &  & \qquad\left.+2c_{2}\frac{p_{x}}{k_{{\rm eff}}}\left\{ F_{x},F_{z}\right\} +c_{2}^{2}\left\{ F_{x},\left\{ F_{y},F_{z}\right\} \right\} \right),\,\label{eq:delta-H-1}
\end{eqnarray}
where we have assumed that $p_{x}/\hbar k_{{\rm _{eff}}}$ and $p_{y}/\hbar k_{{\rm eff}}$
are both $\mathcal{O}\left(1\right)$. At large momenta, $|\mathbf{p}|/\hbar k_{{\rm eff}}\gg1$,
the expansion will break down. For a spin-1/2 system, the anticommutators
vanish, and only the first term proportional to\textbf{ }$p_{x}p_{y}F_{z}$
remains in Eq. (\ref{eq:delta-H-1}). This term and higher order corrections
can also be obtained by expanding, for small pulse durations $\tau$,
the angle $\gamma\left(\mathbf{p}\right)$ featured in the exact Hamiltonian
$H_{2D}^{\left(\mathrm{exact}\right)}$. For higher spin systems,
the anticommutators produce corrections that cannot be expressed using
only the original angular momentum algebra. In general, the $n$-th
order correction will contain nested anti-commutators of the operators
$F_{x}$, $F_{y}$ and $F_{z}$ of order up to $n+2$.

\paragraph*{Interactions:}

We now consider the addition of the interaction Hamiltonian $\mathcal{H}_{I}=\frac{g_{abcd}}{2}\int d\mathbf{r}\psi_{a}^{\dagger}\psi_{b}^{\dagger}\psi_{c}\psi_{d}$,
where $\psi_{a}\left(\mathbf{r}\right)\left(\psi_{a}^{\dagger}\left(\mathbf{r}\right)\right)$
is a Bose or Fermi annihilation(creation) operator for a particle
with spin $a$ at position $\mathbf{r}$, and $g_{abcd}$ is a spin-dependent
interaction constant. The full Hamiltonian in the presence of interactions
is $\mathcal{H}=\mathcal{H}_{0}+\mathcal{H}_{I}$, where $\mathcal{H}_{0}=\int d\mathbf{r}\:\psi_{a}^{\dagger}\left[H_{0}+H_{\mathbf{B}}(\mathbf{r},t)\right]_{ab}\psi_{b}$.
In the absence of an external field to break rotational symmetry,
interactions must be $SU(2)$ invariant, and will be unaffected by
the transformation to the rotating frame which eliminated the magnetic
field. It can be seen that for a sufficiently short pulse, the effective
many-body Hamiltonian is given by $\mathcal{H}_{2D}=\int d\mathbf{r}\psi_{a}^{\dagger}\left(H_{2D}\right)_{ab}\psi_{b}+\mathcal{H}_{I}$.
In other words, the effective spin-orbit coupling does not modify
the form or symmetry of the interactions.

\paragraph*{Experimental implementation}

Quasi-dc magnetic fields $\mathbf{B}({\bf r})$ can be approximated
with the series $B_{j}({\bf r})\approx B_{j}^{(0)}+B_{jk}^{(1)}r_{k}+\cdots$.
Since there are no electrical currents inside the atomic cloud, the
divergence and curl of \textbf{$\mathbf{B}\left(\mathbf{r}\right)$}
are zero. This constrains $B_{jk}^{(1)}$ to be a symmetric traceless
matrix. Hence the magnetic fields used in our previous analysis, such
as ${\bf B}\propto x\ey$ cannot exist and are accompanied by a counter
term such as $y\ex$. The constraint can be lifted by applying a combination
of a strong bias field $B^{(0)}\ez$ and a rf field of frequency $\omega=g_{F}\mu_{{\rm B}}B^{(0)}/\hbar$,
such as ${\bf B}_{x}({\bf r},t)=B^{(0)}\ez+B^{(1)}\left(t\right)\cos(\omega t+\phi_{x})\left(x\ex-z\ez\right)$,
where $B^{(1)}\left(t\right)$ is an envelope function that is slowly
varying compared to $1/\omega$. The bias field and the fast temporal
dependence of the rf field entering $H_{\mathbf{B}}(\mathbf{r},t)$
can be eliminated via a position-independent rotation $S=\exp\left[i\omega tF_{z}/\hbar\right]$
of the spin around $\ez$ with frequency $\omega$. Terms oscillating
at frequencies $\omega$ and $2\omega$ are removed through the rotating
wave approximation in the transformed Hamiltonian $SH_{\mathbf{B}}(\mathbf{r},t)S^{\dagger}-i\hbar S\partial_{t}S^{\dagger}$,
giving Eq. (\ref{eq:H_B}) with 
\begin{align}
\boldsymbol{\Omega}_{x}({\bf r},t) & =\frac{g_{F}\mu_{B}B^{(1)}\left(t\right)}{2\hbar}x\left(\ex\cos\phi_{x}+\ey\sin\phi_{x}\right).\label{eq:b_eff_x}
\end{align}
The field $\boldsymbol{\Omega}_{x}({\bf r},t)=-\beta\left(t\right)k_{{\rm eff}}x\ey$
in the first stage of the 2D setup is obtained with the phase $\phi_{x}=-\pi/2$,
where $\beta\left(t\right)k_{{\rm eff}}=g_{F}\mu_{B}B^{(1)}\left(t\right)/2\hbar$.
Thus the pulsed-gradient magnetic field described in the preceding
sections is represented by the envelope functions which shape the
rf field. 

In the second stage of a two-dimensional setup the magnetic field
${\bf B}_{y}({\bf r},t)=B^{(0)}\ez+B^{(1)}\left(t-\tau\right)\cos(\omega t+\phi_{y})\left(y\ey-z\ez\right)$
leads to 
\begin{align}
\boldsymbol{\Omega}_{y}({\bf r},t) & =\frac{g_{F}\mu_{B}B^{(1)}\left(t-\tau\right)}{2\hbar}y\left(-\ex\sin\phi_{y}+\ey\cos\phi_{y}\right).\label{eq:b_eff-y}
\end{align}
 For $\phi_{y}=-\pi/2$ we reproduce the second stage magnetic field
$\boldsymbol{\Omega}_{y}({\bf r},t)=\beta\left(t-\tau\right)k_{{\rm eff}}x\ey$.
Only the phase difference $\phi_{x}-\phi_{y}$ is relevant, the absolute
phase reflects the choice of the origin of time.

Figure \ref{fig:Experimental-scheme} shows an atom-chip implementation
of the 2D SOC of the Rashba-type. A constant bias field $B^{(0)}\ez$
is applied out of plane, and two pairs of microwires parallel to $\ex$
and $\ey$ provide the rf magnetic fields ${\bf B}_{x}({\bf r},t)$
and ${\bf B}_{y}({\bf r},t)$, respectively. By properly timing the
currents in the pairs of wires, one arrives at the required effective
magnetic couplings $\boldsymbol{\Omega}_{x}({\bf r},t)$ and $\boldsymbol{\Omega}_{y}({\bf r},t)$.
Realistic values \cite{Trinker2008} of $B^{(0)}=20{\rm G}$, $B^{(1)}=0.06{\rm G}/\mu{\rm m}$,
and $\beta\left(t\right)=\frac{2\pi}{\tau}\sin\left(2\pi\frac{t}{\tau}\right)$
with $\tau=50\mu{\rm s}$ give an estimate of $k_{{\rm eff}}\approx1\mu{\rm m}^{-1}$,
compared to optically induced SOC in Rubidium where $k_{{\rm eff}}\approx8\mu{\rm m}^{-1}$
\cite{Lin2011,Chen2012}. The creation of a 3D SOC would be a much
more challenging experimental task. In that case one not only needs
to use several rf pulses with the magnetic field oriented along different
planes, but also periodically alter the direction of the bias field.

\paragraph*{Summary:}

We proposed a scheme to simulate Rashba spin-orbit coupling in an
arbitrary spin-$f$ gas of ultracold atoms. The scheme used pulsed
magnetic field gradients along perpendicular directions to impart
a spin-dependent momentum boost to the atoms. For sufficiently short
evolution time, the time-averaged Hamiltonian well approximated the
Rashba Hamiltonian. Higher order corrections to the energy spectrum
were calculated exactly for spin-$1/2$ and perturbatively for higher
spins. We then considered interactions, and found that for short pulses,
the form of the interactions is not modified. Finally, we proposed
an experimental implementation of such a scheme on atom-chips.

\paragraph{Note:}

After submission of this manuscript, an article by Xu, et. al., \cite{Xu2013}
appeared that also considers Rashba SOC using pulsed magnetic field
gradients.

\paragraph*{Acknowledgements:}

This work was initiated at the Nordita workshop ``Pushing the Boundaries
of Cold Atoms''. G.J. acknowledges the financial support by the Lithuanian
Research Council Project No. MIP-082/2012. I.B.S. and B.M.A. acknowledge
the financial support by the NSF through the Physics Frontier Center
at JQI, and the ARO with funds from both the Atomtronics MURI and
DARPA's OLE Program. Helpful discussions with\textbf{ }B. Blakie,
A. Eckardt, M. Foss-Feig, S.-C. Gou, H. Pu, J. Ruseckas, L. Santos,
V. Shenoy, U. Schneider and R. Wilson are gratefully acknowledged.

\bibliographystyle{apsrev}
\bibliography{Rashba-without-Light}

\end{document}